%% file: radgal.tex
\documentstyle[itzidef]{mn}
\input epsf

\title[Stellar populations in the nuclear regions of nearby 
radiogalaxies]
{Stellar populations in the nuclear regions of nearby radiogalaxies}

\author[I. Aretxaga, E. Terlevich, R. Terlevich, G. Cotter, 
A.I. D\'{\i}az]{Itziar Aretxaga$^1$, 
Elena Terlevich$^{1}$\thanks{Visiting Fellow at
IoA, UK}, 
Roberto J. Terlevich$^{2}$\thanks{Visiting Professor at 
INAOE, Mexico}, Garret Cotter$^3$, 
\\ \\ 
{\LARGE \'Angeles I. D\'{\i}az$^{4}$} \\
$^1$ Instituto Nacional de Astrof\'{\i}sica, \'Optica y 
Electr\'onica, Apdo. Postal 25 y 216, 72000 Puebla, Pue., Mexico\\
$^2$ Institute of Astronomy, Madingley Road, Cambridge CB3 0HA, U.K.\\
$^3$ Cavendish Laboratory, Univ. of Cambridge, Madingley Road, 
Cambridge CB3 0HE, U.K.\\
$^4$ Dept. F\'{\i}sica Te\'orica C-XI, Univ. Aut\'omoma de Madrid, 
Cantoblanco, Madrid, Spain.\\
}

\begin{document}

\maketitle

\begin{abstract}
We present optical spectra of the nuclei of seven luminous 
($P_{\mbox{\scriptsize 178MHz}} \gsim 10^{25}$~W~Hz$^{-1}$~Sr$^{-1}$)
nearby ($z<0.08$) radiogalaxies, which 
mostly correspond to the FR~II class. 
In two cases,
Hydra~A and 3C~285, the Balmer and \ldo{4000} break indices 
constrain the spectral types and luminosity classes
of the stars involved, revealing that the blue spectra are dominated
by blue supergiant and/or giant stars. The ages derived for the last burst 
of star formation in Hydra~A are between 7 and 40~Myr, and in 3C~285
about 10~Myr. The rest of the narrow-line radiogalaxies (four) have 
\ldo{4000} break and metallic indices consistent with those of elliptical 
galaxies. The only broad-line radiogalaxy
in our sample, 3C~382, has a strong featureless blue continuum
and broad emission lines that dilute the underlying blue stellar spectra.
We are able to detect the Ca II triplet 
in absorption in the seven
objects, with good quality data for 
%\end{abstract} {\small
only four of them.
The strengths of the absorptions are similar to 
those found in normal elliptical galaxies, but these values are both
consistent with single stellar populations of ages as derived from 
the Balmer 
absorption and break strengths, and, also, with mixed young$+$old populations.
%}
\end{abstract}

\begin{keywords}
galaxies: active -- galaxies: starbursts -- galaxies: stellar content
\end{keywords}

%%%%%%%%%%%%%%%%%
%%%% Introduction
%%%%%%%%%%%%%%%%%

\section{Introduction}

In recent years new 
evidence that {\it star formation plays an important role in Active
Galactic Nuclei} (AGN) has been gathered:
\begin{description}
	\item[$\bullet$]
 The presence of strong  
Ca~II~$\lambda\lambda$8494,8542,8662\AA\ triplet 
(CaT) absorptions in a large sample of Seyfert~2 nuclei has provided
direct evidence for a population of red supergiant stars that dominates
the near-IR light (Terlevich, D\'{\i}az
\& Terlevich 1990). The values found in Seyfert~1 nuclei are
also consistent with this idea if the dilution produced 
by a nuclear non-stellar 
source is taken into account (Terlevich, D\'{\i}az
\& Terlevich 1990, Jim\'enez-Benito et al. 2000).
The high 
mass-to-light ratios $L(1.6 \mu \mbox{m})/M$ inferred in Seyfert~2 nuclei
also indicate that red supergiants dominate the nuclear light 
(Oliva et al. 1995), but a similar conclusion does not hold 
for Seyfert~1 nuclei.

	\item[$\bullet$]
 The absence of broad emission lines in the direct optical spectra
of Seyfert~2 nuclei which show broad lines in polarized light can be understood
only if there is an additional central source of continuum, most probably blue
stars (Cid Fernandes \& Terlevich 1995, Heckman et al. 1995).  
This conclusion is further supported by the detection of 
polarization levels which are lower in the continuum than in the broad lines
(Miller \& Goodrich 1990, Tran, Miller \& Kay 1992).
	\item[$\bullet$]
Hubble Space Telescope imaging of the 
Seyfert Mrk~447 reveals that the central UV
light arises in a {\em resolved} region of a few hundred pc, in which
prominent CaT absorption and broad He~II~\ldo{4686} emission lines
reveal the red supergiant and Wolf Rayet stars of a powerful
starburst. The stars dominate the UV to near-IR light directly 
received from the nucleus (Heckman et al. 1997). At least 50~per cent of the
light emitted by the nucleus is stellar,
as a conservative estimate.
Mrk~447 is not a rare case: a large
sample of nearby bright Seyfert~2s and LINERs show similar 
resolved  starburst nuclei of 80 to a few hundred pc in size 
(Colina et al. 1997, Gonz\'alez-Delgado et al. 1998, Maoz et al. 1995, 1998),
with some of the Seyfert~2 containing dominant Wolf-Rayet populations 
(Kunth \& Contini 1999, Cid Fernandes et al. 1999).

\end{description}

A starburst--AGN connection has been proposed in at least three
scenarios: starbursts giving birth to massive black holes
(e.g. Scoville \& Norman 1988); 
black holes being fed by surrounding stellar clusters (e.g. 
Perry \& Dyson 1985, 
Peterson 1992); and
also pure starbursts without black holes (e.g. Terlevich \& Melnick 1985,
Terlevich et al. 1992). The evidence for starbursts in Seyfert nuclei
strongly supports some kind of connection. However, it is still to be
demonstrated that starbursts play a key role in {\em all} kinds of 
 AGN.

One of the most stringent tests {\em to assess if all AGN
have associated enhanced nuclear star formation} is the case of 
lobe-dominated radio-sources, whose host galaxies have relatively red colours
when compared to other AGN varieties. 
In this paper we address the stellar content 
associated with the active nuclei of a sample of FR~II radiogalaxies, 
the most
luminous class of radiogalaxies (Fanaroff \& Riley 1974)
which possess the most powerful central
engines and radio-jets (Rawlings \& Saunders 1991). 
The presence of extended
collimated radio-jets, which fuel the extended radio structure over
$\gsim 10^8$~yr, strongly suggests the
existence of a supermassive accreting black hole in the nuclei of these 
radiogalaxies.
 This test addresses the question 
of whether AGN that involve conspicuous black holes and accretion 
processes also contain enhanced star formation.

In section~2 we introduce the sample and detail the data acquisition and 
reduction processes. In section~3 we provide continuum and line
measurements of the most prominent features of the optical spectra
of the radiogalaxies. In section~4 we discuss the main stellar
populations responsible for the absorption and continuum spectra.
In section~5 we offer notes on individual objects. A sumary of the main 
conclussions from this work is presented
in section~6.

%%%%%%%%%%%%%%
%%%% Section 2
%%%%%%%%%%%%%%

\ifoldfss
  \section{Data acquisition and reduction}
\else
  \section[]{Data acquisition and reduction}
\fi

Our sample of radiogalaxies was extracted from the 3CRR catalogue
(Laing, Riley and Longair 1983) with the only selection criteria being
edge-brightened morphology, which defines the
FR~II class of radiogalaxies (Fanaroff \& Riley 1974),
and redshift $z < 0.08$.  This last condition 
was imposed in order to be able to observe the redshifted CaT at wavelengths
shorter than \ldo{9300}, where the 
atmospheric bands are prominent.
Six out of a complete sample of ten FR~II radiogalaxies that fulfill these
requirements were
randomly chosen. In addition to this sub-sample
of FR~IIs, 
we observed the unusually luminous FR~I radiogalaxy Hydra A (3C~218). This has 
a radio luminosity of \mbox{$P_{178{\rm MHz}} = 2.2 \times 10^{26}$ 
W~Hz$^{-1}$~Sr$^{-1}$}, which is an order of magnitude above the typical 
FR~I/FR~II dividing luminosity.

%The sample we study in this paper was extracted from
%the 3C catalog (Laing, Riley \& Longair 1983) with the only
%selection criteria of having a radio-power above 
%$P_{1\mbox{\scriptsize 1GHz}} \sim 10^{25.5}$~W~Hz$^{-1}$, which 
%characterizes the FR~II class of radiogalaxies (Fanaroff \& Riley 1974),
%and being at redshifts $z<0.08$.  The seven radiogalaxies we
%study in this paper form a random sample out of the ten 3C galaxies 
%which fulfill these two requirements.

Spectroscopic observations of a total of seven radiogalaxies, one normal
elliptical galaxy to serve as reference and
five K~III stars to serve as velocity calibrators 
were performed using the double-arm 
spectrograph ISIS mounted in the
Cassegrain Focus of the 4.2m William Herschel Telescope\footnote{The William Herschel Telescope 
is operated on the island of La Palma by the Isaac Newton Group
in the Spanish Observatorio del Roque de los
Muchachos of the Instituto de Astrof\'{\i}sica de Canarias} in La
Palma during two observing runs, in 1997 November 7--8 and 1998 February 
19--20.  The first run was photometric but the second
was not, being partially 
cloudy on the 20th. The seeing, as measured from the spatial
dimension of spectrophotometric stars,  was between $0.7$ and $0.8$~arcsec
throughout the nights. 

A slit width of 1.2~arcsec centered on the nucleus of
galaxies and stars was used. 
We oriented the slit along the radio-axis 
for all radiogalaxies, except for Hydra~A,
for which the orientation was perpendicular to the radio-axis.

An R300B grating centered at \ldo{4500} with a 2148x4200 pixel EEV CCD
and an R316R grating centered at \ldo{8800} with a 1024x1024 pixel TEK CCD
were used in the 1998 run.  
The projected area on these chips is 0.2~arcsec/pixel and 0.36~arcsec/pixel 
respectively.
This configuration provides the spectral resolution necessary to resolve
the Mg~b and CaT features and, at the same time, offers a wide spectral span:
\ldo{3350}---\ldo{6000} at 5.1\AA\ 
resolution in the blue and 
\ldo{7900}---\ldo{9400} at 3.5\AA\ resolution in the red. 
In the 1997 run, in which we assessed the viability 
of the project, we used the R600B and R600R  
gratings instead. This setup covers the \ldo{3810}---\ldo{5420} and
\ldo{8510}---\ldo{9320} range in the blue and red arm, at 2.6 
and 1.7\AA\ resolution respectively.
Just one radiogalaxy (DA~240) was observed with this
alternative setup. 
The dichroics 5700  and 6100  were used in 1997 November
and 1998 February, and in both runs we used a filter
to avoid second order contamination in the spectra.

We obtained flux standards (HZ~44 and G191-B2B) for the four nights 
and gratings, except in 1998 
Feb 20, when we were unable to acquire the red spectrum of the
corresponding standard due to a technical failure.
One calibration 
lamp CuAr$+$CuNe exposure per spectral region and telescope position was also 
obtained for all objects. 

The total integration times for the radiogalaxies (from 1 to 3 hr)
were split into time intervals of about 1200 or 1800~s in order
to diminish the effect of cosmic rays on individual frames and
allow to take  lamp flat-fields with  the red arm of the spectrograph 
between science exposures. 
The TEK CCD  
has a variable fringing pattern at the wavelengths of interest,
such that the variations are correlated  with the position at which the 
telescope is pointing. Since flat-fielding is crucial for the 
reddest wavelengths, where the sky lines are most prominent, 
after every exposure of
20 to 30 min we acquired a flat-field in the same position 
of the telescope as the one for which the galaxies were being observed. 
We followed this procedure with all galaxies except with DA~240.
The same procedure was also used in the case of
the elliptical galaxy, splitting its total integration time in two.

Table~\ref{obs} summarizes the journal of observations, where
column~1 gives the name of the object; column~2 the radio-power at
178~MHz; column~3 the redshift; 
column~4 the integrated
$V$ magnitude of the galaxy; column~5 identifies
whether the object is a radiogalaxy (RG), a normal elliptical (E) 
or a star (S);
column~6 gives the date of the beginning of the night in which the 
observations were carried out; column~7 the position angle (PA) of the slit;
column~8 the total exposure time; column~9 the grating used; and column~10
the corresponding linear size to 1~arcsec at the redshift of the galaxies
  (for \Ho50).
The data for the radiogalaxies were extracted from the 3C Atlas (Leahy, 
Bridle \& Strom, {\tt http://www.jb.man.ac.uk/atlas/}) and for the host 
galaxy of Hydra~A from the 
3CR Catalogue (de Vaucouleurs et al. 1991).

The data were reduced using the IRAF software package. 
The frames were first bias subtracted and then 
flat-field corrected. In the case of the red arm spectra, the different 
flats obtained for a single object were combined 
when no significant differences were detected between them. However, in 
several cases
the fringing pattern shifted positions that accounted for differences of up to 
20~per cent. In these cases we corrected each science frame with the 
flat-field acquired immediately before and/or afterwards. 
Close inspection of the
faintest levels of the flat-fielded frames showed that the fringing had been 
successfully eliminated. Wavelength and flux calibration
were then performed, and 
the sky was subtracted by using the outermost parts of the slit.

The atmospheric bands, mainly water absorption at \ldo{8920}---\ldo{9400},
affect the redshifted CaT region of several radiogalaxies. 
The bands have been extracted
using a template constructed from the stellar spectra
obtained each night. The template was 
built averaging the normalized flux of spectrophotometric 
and velocity standard stars, once 
the stellar absorption lines had been removed. The atmospheric
bands were eliminated 
from the spectra of the galaxies dividing by the flux-scaled template. 
This reduces the S/N
of the region under consideration, especially since the bands are variable 
in time and one of our observing nights was partially cloudy.
However, the technique  allows the detection of the stellar 
atmospheric features. The CaT of the elliptical galaxy is not affected 
by atmospheric absorption.

  Figure~1 shows the line spectrum of the sky and, as an example, 
the atmospheric absorption template of 1998 Feb 19.
Water-band
correction proved to be critical for the detection
of the CaT lines when the atmospheric conditions were most adverse.

%  Figure~2 shows grey plots of the spectra of the seven radiogalaxies
%and the elliptical galaxy.

  Figure~2 shows extractions of the nuclear 2~arcsec of 
the spectra of the galaxies. This corresponds to
844 to 2020~pc for the  radiogalaxies, and 98~pc
for the normal elliptical galaxy.

%%%%%%%%%%%%%%
%%%% Section 3
%%%%%%%%%%%%%%

\ifoldfss
  \section{Line and continuum measurements}
\else
  \section[]{Line and continuum measurements}
\fi

\subsection{CaT index}

The CaT was 
detected in all of the objects, although in three cases (3C~285, 3C~382 and 
4C~73.08) it was totally or partially affected by residuals left by
the atmospheric band corrections 
and the measurement of its strength was thus precluded. 
For the remaining cases, the strength was measured in the rest-frame of 
the galaxies 
against a pseudo-continuum, following the definition of the CaT index
of D\'{\i}az, Terlevich \& Terlevich (1989). 
In Hydra~A,
3C~285 and 3C~382, the red continuum band is seriously 
affected by residuals left
from the atmospheric absorption removal. We defined two 
alternative continuum bands, \mbox{ 8575 \AA  $< \lambda <$ 8585 \AA} and 
\mbox{8730 \AA $< \lambda <$ 8740 \AA}, that substitute the red-most continuum
window of the CaT index. We checked this new definition against the
original one in the elliptical galaxy, which doesn't have 
residuals in its continuum bands, and the 
agreement between the two systems was good within 5~per cent.

Velocity dispersions were measured by cross-correlating 
the galaxy spectra with the stellar spectra obtained with the same
setup. The errors in the velocity dispersions calculated in this way
were less than 10~per cent.

A high velocity dispersion tends to decrease the measured values of indices
based on EW measurements. The CaT index 
has to be corrected from broadening of the absorption lines 
by the corresponding velocity dispersion.
In order to calculate the correction we convolved stellar profiles with 
gaussian functions of increasing width, and measured the CaT index in them.
A good description of the correction found for our data is given by the 
functional form \mbox{$\Delta$EW (\AA ) $=(\sigma(\mbox{\univel}) -100)/200$}.
The corrections were applied to the values measured 
in the galaxies, and converted into unbroadened indices.

The values of velocity dispersions ($\sigma$), uncorrected EW 
(CaT$_{\mbox{\scriptsize u}}$) and corrected EW (CaT), 
are listed in Table~\ref{EW}.

\subsection{\ldo{4000} and Balmer Break indices}

Stellar populations can be dated through
the measurement of the \ldo{4000} or Balmer breaks. 
In intermediate to old populations the discontinuity at \ldo{4000} results 
from a combination of the accumulation of the 
Balmer lines towards the limit  of the Balmer absorption 
continuum  at \ldo{3646} (the Balmer break) and the increase 
in stellar opacity caused by
metal lines shortwards of \ldo{4000}. 

     Table~3 lists the values of the \ldo{4000} break index, $\Delta$4000\AA,
measured in the spectra of the 6 narrow-line radiogalaxies and the elliptical 
galaxy in our sample. This excludes 3C~382, which has a spectrum dominated 
by a strong blue continuum and broad-emission lines, and shows very weak 
stellar atmospheric features and no break.
We adopted 
the definition given by Hamilton (1985), which quantifies the ratio of the 
average flux-level of two broad bands, one covering the break 
(3750-3950) and one bluewards of the break (4050-4250). Both bands contain
strong metallic and Balmer absorption lines in the case of normal galaxies.
In active galaxies, the measurement can be contaminated by emission 
of [Ne III]\ldo{3869}, which in our case is weak. The contamination by
high-order Balmer lines in emission is negligible. The net effect of emission
contamination is to decrease the Balmer break index. In the radio-galaxies, 
we have estimated this effect by interpolating the continuum levels below the
[Ne III] emission, and we estimate that the ratio can be affected by 
6~per cent at worst, in the case of 3C~192, and by less than 3~per cent for the
rest of the objects. Table~3 lists emission-devoid indices.

  Hydra~A and 3C~285 have spectra which 
are much bluer than those of normal elliptical galaxies. 
In order to  quantify better
the strength of the break and the ages of the populations derived, we have 
performed a bulge subtraction using as template the spectrum of NGC~4374, 
scaled to eliminate the G-band absorption 
of the radiogalaxies. Since the velocity dispersion of the stars in NGC~4374
and in the radiogalaxies are comparable inside the spectral resolution of our 
data, no further corrections were needed. The G-band absorption is prominent 
in stars 
of spectral types later than F5 and it is especially strong in types K. 
NGC~4374 is a normal elliptical galaxy, with a spectral shape which
compares well with those of other normal ellipticals in the spectrophotometric
atlas of galaxies of Kennicutt (1992).
Thus, by removing a scaled template of NGC~4374, we are isolating
the most massive stars ($M \gsim 1$\Msun)
in the composite stellar population of the radiogalaxies.
Figure~3 shows the bulge subtractions obtained on these
two radiogalaxies. 

We measured on the bulge-subtracted spectra $\Delta$4000\AA\
and also the Balmer break index as defined by the
classical $D\lambda_1$ method of stellar classification designed by Barbier
and Chalonge (Barbier 1955, Chalonge 1956, see Str\"omgren 1963).
The latter quantifies the Balmer discontinuity in terms of 
 the logarithmic difference
of the continuum levels ($D$) and the effective position of the break
($\lambda_1$). The method places a pseudo-continuum on top of the
higher order terms of the Balmer series in order to measure the effective 
position of the discontinuity. Figure~4 shows the placement of continua, 
pseudo-continua and the measurements of
$D$ and $\lambda_1$ for an A2I star from the stellar library of 
Jacoby, Hunter \& Christian (1984). The functional dependences
on the effective temperature and gravity of the stars are sufficiently
different for $D$ and $\lambda_1$ to satisfy a two-dimensional
classification.

The $D\lambda_1$ method could only be reliably applied 
in the cases of Hydra~A and 3C~285. 
For the other radiogalaxies, the bulge-subtractions led to results 
that did not allow the identification of the absorption features
and/or the break in an unambiguous way due to the resulting 
poor S/N.
Figure~5 shows the $D\lambda_1$ measurements performed on the 
bulge-subtracted spectra of Hydra~A and 3C~285. We have placed different 
continuum levels to estimate the maximum range of acceptable
parameters of the stellar populations that are involved.

Table~\ref{break} lists the \ldo{4000} and Balmer break indices measured in 
both the bulge-subtracted and the original spectra of the
radiogalaxies.

\subsection{Lick indices}

The presence of  prominent Balmer absorption lines, from 
H$\gamma$ up to H12~\ldo{3750}, is
one of the most remarkable features of the blue spectra of two of the seven 
radiogalaxies,
while   H$\beta$ and H$\alpha$
are filled up by conspicuous emission lines.
A clear exception to the presence of the Balmer series in absorption is 
the broad-line radiogalaxy 3C~382.

  In order to estimate the Balmer strength, crucial to date 
any young stellar population involved, we use the EW
of the H10~\ldo{3798} line, 
which appears only weakly contaminated by emission
in the radiogalaxies. 
H10 is chosen as a compromise of an easily detectable
Balmer line that shows both a minimum of emission contamination and clear wings
to measure the adjacent continuum. The Balmer lines from \Hb\ to 
H9~\ldo{3836} are contaminated by prominent emission, which in Case~B 
recombination comes in decreasing emission ratios to \Hb\ 
of 1, 0.458, 0.251, 0.154, 0.102, 0.0709
(Osterbrock 1989);  H10 has an emission contamination of $0.0515 \times \Hb$.
At the same time, the absorption strengths are quite similar
from \Hb\ to H10, although
the EW(H10) is actually systematically smaller than 
EW(\Hb) in young to intermediate-age populations. 
Gonz\'alez-Delgado, Leitherer \& Heckman  
(2000) obtain, in their population synthesis models, ratios of EW(\Hb )/EW(H10)
between 1.3 and  1.6 for bursts with ages 0 to 1 Gyr and constant or coeval 
star formation histories. 
Lines of order higher than 10
have decreasing emission contamination, but they also
increasingly merge towards the Balmer continuum limit. 

A caveat in the use of H10 as an age calibrator comes from the
fact that this line might be contaminated by metallic lines in old populations.
Although our measurements of H10 in NGC~4374 are around 1.5\AA, an
inspection of the spectra of three elliptical galaxies (NGC~584, NC~720,
NGC~821) observed in the 
same wavelegth range (but with lower S/N) and archived in the 
Isaac Newton Group database,
indicates that a wide range of EW(H10), from 2 to 4\AA, 
could characterize elliptical galaxies, while their
\Hb\ indices are in the 1 to 2\AA\ regime. If confirmed by better data,
these results could indicate that although the upper Balmer series is detected
in elliptical galaxies, it could indeed be contaminated by the absorptions of
other species. Clearly, more
work needs to be done in the near-UV spectra of elliptical galaxies
before conclusive evidence
can be derived for the behaviour of EW(H10) in old stellar systems, and its
contamination by metallic lines.

In all the radio-galaxies observed in this work, the H10 profile 
is narrow and 
reproduces the shape of the wings of the lower-order Balmer absorption lines.
Hydra~A and 3C~285 clearly provide the best fittings.
As an illustration, Figure~6 shows the estimated absorption line profiles 
for the H$\beta$, H$\gamma$
and H$\delta$ lines, assuming a constant ratio between their 
EWs  and that of H10,
and also a scaled ($\times 1.4$) H10 profile for the case of \Hb\ in Hydra~A.

We also measured indices that are mostly sensitive to
the metal content of the stellar populations involved. The Lick
indices of Mg and Fe (e.g. Worthey et al. 1994) serve this purpose. 
In order to avoid the contribution of \mbox{[O III]\ldo{4959}} 
to the continuum measurement for the molecular index Mg$_2$, 
we have displaced
the lower continuum band of this index to \mbox{4895.125 \AA $< \lambda < $
4935.625 \AA}.
This redefinition does not alter the value of the index in the 
elliptical galaxy, which shows no [O III] emission.

Table~\ref{lick} compiles the EW of H10,  and the metallic indices 
Mg b,  Fe5270, Fe5335, [MgFe], Mg$_2$  of the Lick system, measured in the 
rest-frame of the galaxies in our sample.
The atomic indices are affected by broadening, like the CaT index, 
while Mg$_2$ is only affected by lamp contributions in the original IDS 
Lick system (Worthey et al. 1994, Longhetti et al. 1998). We have calculated 
broadening corrections as in section 3.1 for the atomic lines,
and adopted the corrections calculated by Longhetti (1998) for the 
molecular lines. The uncorrected 
values of these indices
are denoted with a subindex ${\mbox{\small u}}$ in 
Table~\ref{lick}. The errors of the individual line and molecular indices were
estimated adopting continua shifted from the best fit continua by $\pm 1\sigma$.
This lead to average errors between
an 8 and a 10~per cent for individual line and molecular indices, and 
$\sim 6$~per cent for [MgFe].

The agreement between our measurements of Lick indices 
and those carried out by other
authors (Gonz\'alez 1993,
Davies et al. 1987, Trager et al. 2000a) on our galaxy in common, NGC~4374, 
is better than 10~per cent.

%%%%%%%%%%%%%%
%%%% Section 3
%%%%%%%%%%%%%%

\ifoldfss
  \section{Discussion}
\else
  \section[]{Discussion}
\fi

\subsection{Comparison with elliptical galaxies and population synthesis
models}

The analysis of the spectral energy distributions and colours of
elliptical galaxies suffers from a well known age-metallicity
degeneracy (Aaronson et al. 1978). However, this is broken down when
the strengths of suitable stellar absorption lines are taken into account
(e.g.~Heckman 1980).
The plane composed by the [\Hb ] and [MgFe] indices, in this sense,
can discriminate the ages and metallicities of stellar systems. It is on the
basis of this plot, that a large spread of ages in elliptical galaxies
has been suggested (Gonz\'alez 1993). Bressan,
Chiosi \& Tantalo (1996) claim, however, that when the UV emission
and velocity dispersion of the galaxies
are taken into account, the data are only compatible with basically old
systems that have experienced different star formation histories
(see also Trager et al. 2000a, 2000b).
A recent burst of star formation that involves only a tiny fraction
of the whole elliptical mass in stars, would rise the [\Hb ] index 
to values characteristic of single stellar populations which 
are 1 to 2 Gyr old (Bressan et al. 1996).

Most likely, the stellar populations of radiogalaxies are also
the combination of
different generations.
Direct support for this interpretation in the case of Hydra~A
comes from the fact that the stellar populations responsible for the 
strong Balmer lines are dynamically decoupled from those
responsible for the metallic lines (Melnick, Gopal-Krishna \& Terlevich 1997).

This interpretation
is also consistent with
the modest $\Delta$4000\AA\  measurements we have obtained.
Figure~7 shows a comparison of the values found in radiogalaxies,
with those of normal elliptical, spiral and irregular galaxies,
including starbursts, from the atlas of Kennicutt (1992).
The radiogalaxies
3C~98, 3C~192, 4C~73.08 and DA~240 have indices of the order
of 1.9 to 2.3, which overlap with those of normal E galaxies,
\mbox{$\Delta$4000\AA\ $=2.08 \pm 0.23$}.
These values 
correspond to populations dominated by stars
of ages 1 to 10 Gyr old, if one assumes the coeval population 
synthesis models of Longhetti et al. (1999). 
However, Hydra~A and 3C~285 have indices in the range
1.4 to 1.6, typical of coeval populations which are 200 to 
500~Myr old.
Once the bulge population is subtracted,
the $\Delta$4000\AA\ indices of Hydra~A and 3C~285 decrease to 1.2 and 1.0 
respectively, which are typical of systems younger than about 60~Myr.

Hamilton (1985) measured the $\Delta$4000\AA\ index in a sample of stars 
covering a wide range of spectral types and luminosity classes. He found 
a sequence of increasing $\Delta$4000\AA\ from B0 to M5 stars, with values  
from 1 to 4~mag respectively. A comparison with the sequence 
he found leads us to conclude
that  the break in the bulge subtracted spectrum of Hydra~A is dominated by 
B or earlier type stars while that of 3C~285 is dominated by A type stars. 
The index $\Delta$4000\AA\ does not clearly  discriminate luminosity classes 
for stars with spectral types earlier than G0.

The equivalent width of the H10 absorption line in these two radiogalaxies 
give further support to the
interpretation of the Balmer break as produced by a young stellar population.
In Hydra~A we find after bulge subtraction EW(H10)$\approx 3.9$\AA,
which, according to the synthesis models
of Gonz\'alez-Delgado et al. (2000) gives ages of 
7 to 15~Myr for an instantaneous burst of star formation, and
40 to 60~Myr for a continuous star formation mode, in solar 
metallicity environments.
In the case of 3C~285, 
EW(H10)$\approx 6$\AA\ would imply an age older than about 
25~Myr for a single-population burst of solar metallicity.

The metallic indices of normal elliptical galaxies range  between the values 
$0.56 \lsim \log {\rm [MgFe]} \lsim 0.66$ (Gonz\'alez 1993), which 
characterizes
oversolar metallicites for ages larger than about 5~Gyr. 
This is also the typical range of our radiogalaxies, 
although 3C~285 shows a clear departure with $\log {\rm [MgFe]} \approx 0.4$.
However, [MgFe] tends to be smaller for populations younger than a few
Gyr and similar
oversolar metallic content. Since 3C~285 has a clear burst of recent 
star-formation, we conclude that its overall abundance is also most probably
solar or oversolar.

\subsection{The blue stellar content}

A better estimate of the spectral type and luminosity class of the
stars that dominate the break in Hydra~A and 3C~285 comes from the 
two-dimensional classification of Barbier and Chalonge.
In Figure~8 the solid squares connected by lines represent the maximum range 
of possible $D\lambda_1$ values measured in these radiogalaxies. 

The
Balmer break index is sensitive to the positioning of the pseudo-continuum
on top of the higher order Balmer series lines, which in turn is
sensitive to the merging of the wings of the lines,
enhanced at large velocity dispersions. In order to assign spectral types and 
luminosity classes to the stars that dominate the break, therefore, it is not 
sufficient to
compare the values we have obtained with those measured in stellar 
catalogues. The values measured for the radiogalaxies can be corrected for
their intrinsic velocity dispersions; we have chosen instead 
to recalibrate
the index using 
 template stars of different spectral types and 
luminosity classes
convolved 
with gaussian functions, until they reproduce the width of the Balmer lines 
observed in the radiogalaxies (FWHM$\approx 12.5$\AA).
We used the B0 to A7 stars
from  the  stellar library of Jacoby et al. (1984), 
which were observed with 4.5\AA\
resolution.  
The values of
the $D\lambda_1$ indices measured in these broadened stars
are represented in Figure~8 by their respective
classification. By comparison we also plot the grid traced by the locus of
unbroadened stars, as published by Str\"omgren (1963).
The broadening of the lines shifts the original locus of 
supergiant stars from the $\lambda_1 \lsim 3720$\AA\
range (Chalonge 1956)  to the $3720 \lsim \lambda_1 \lsim 3740$\AA\ range, 
occupied by giant stars in the original (unbroadened) classification. 
Giant stars, in turn, shift to
positions first occupied by dwarfs. Most dwarfs have Balmer 
line widths comparable to those of the radiogalaxies, and thus their locus 
in the 
diagram is mostly unchanged.

The value of the $D$ index indicates that the recent burst in 
Hydra~A is dominated by  B3 to B5 stars, and the
effective position of the Balmer break ($\lambda_1$) indicates that
these are giant or supergiant stars, respectively. 
These stars have masses of
7 and 20\Msun\ (Schmidt-Kaler 1982). From the stellar 
evolutionary tracks of massive stars with standard mass-loss rate
at \Zsun\ or 2\Zsun\
(Schaller et al. 1992, Schaerer et al. 1993, Meynet et al. 1994) we infer 
that these stars must have ages
between 7 to 8 Myr (B3I) and 40 Myr (B5III). 
Note that the B4V stars in Figure~8, near the location of Hydra~A, cannot
originate the break and at the same time follow the kinematics of the
nucleus (see section 5.3).
Any dwarf star located in the 
stellar disk of Hydra~A
would show
absorption lines that have been broadened beyond
the 12.5\AA\ of FWHM we measure in
this radiogalaxy, and its position would have been shifted further into
larger values of $\lambda_1$.

The location in the $D-\lambda_1$ plane of 3C~285 indicates that the break is
produced by A2I stars. These are 15\Msun\ stars. Again,
ages of 10 to 12~Myr  are found for the last burst of star formation 
in this radiogalaxy. The interpretation of the blue excess in terms
of A type stars is further supported by the detection of the Ca~II~H  line
in the bulge-subtracted spectrum.

\subsection{The red stellar content}

The CaT index in the radiogalaxies has values between 6 and 7\AA.
D\'{\i}az et al. (1989) 
find that at solar or oversolar metallicities red supergiant stars  
have CaT indices ranging from 8.5 to 13\AA, red giant stars from 6 
to 9\AA\ and
dwarfs from 4.5 to 8.5\AA. The values we find are thus compatible with 
both giant or dwarf stars. However, we favour the interpretation of
giant stars since the old bulge population will be dominated by red giants,
as in the case of elliptical galaxies. 

We have measured the CaT
in a control sample of elliptical galaxies observed by J. Gorgas and 
collaborators (priv. communication) and combined these 
measurements with those found by Terlevich et al (1990) in 
a sample of elliptical,
lenticular, spiral, and active galaxies of different kinds. 
We find that the range of CaT
in elliptical galaxies, 5 to 7.5\AA , comprises the range of values of the
radiogalaxies (see Figure~9). 

Garc\'{\i}a-Vargas,
Moll\'a \& Bressan (1998) find in their population synthesis models
values of the CaT between 6 and 7\AA\ for ages ranging from
100 Myr to 1 Gyr, and larger afterwards, assuming coeval star formation 
and solar or oversolar metallicity.  A revised version of these models 
by Moll\'a \& Garc\'{\i}a-Vargas (2000) which includes extended libraries of 
M-type stars, predicts for ages between 2 and 20~Gyr a constant value 
of 7\AA\ at solar metallicity, and 8.5\AA\ at 2\Zsun.
These synthesis  models are based on parametric fittings 
of the CaT strength in NLTE model atmospheres 
(J\o rgensen,  Carlsson \& Johnson
1992) and in fittings of empirical values  measured
in stellar libraries (e.g. D\'{\i}az et al. 1989; Zhou 1991). 
The fits work well in the low metallicity regime. However,
at metallicities higher than solar, the relationship between metallicity and
the CaT index shows a big scatter, and the linear fittings loose any 
predictive power
(see Figure~4 of D\'{\i}az et al. 1989). 

Red supergiant stars appear in coeval population synthesis models
between 5 and 30 Myr, and create
a maximum strength of the CaT index (CaT $\gsim 9$\AA) around
6 to 16 Myr for \Zsun\ and 5 to 30 Myr for 2\Zsun\
(Garc\'{\i}a-Vargas et al. 1993, 1998; Mayya 1997).
Strengths 
of CaT $\gsim 7$\AA\ are characteristic of
bursts with ages between 5 and 40~Myr.
Leitherer et al. (1999) also find that the total strength of the population
depends dramatically on the slope of the initial mass function (IMF) 
and star formation history. 
While a coeval burst with a complete  Salpeter IMF yields values 
surpassing 7\AA\ between 
6 and 12~Myr, the same IMF in a continuous star formation
mode yields values of only 5.5\AA\ maximum. The CaT 
strength values for coeval star formation derived by Leitherer et al. (1999) 
differ substantially from those derived by Garc\'{\i}a-Vargas and
coworkers (1993, 1998) and by Mayya (1997), most probably due to a different
calibration of the CaT index.

Mixed populations of young bursts which contain red supergiants,
superposed on old populations can
also yield values of the CaT between 4 and 8\AA\ (Garc\'{\i}a-Vargas 
et al. 1998). Since metal rich giant stars  
have CaT values ranging from 6 to 9\AA\, we regard our
observations of the CaT index in radiogalaxies, as being compatible with 
ages 1 to 15 Gyr.

%%%%%%%%%%%%%%
%%%% Section 5
%%%%%%%%%%%%%%

\ifoldfss
  \section{Notes on individual objects}
\else
  \section[]{Notes on individual objects}
\fi

\subsection{3C 98}

3C~98 shows a double-lobe radio structure which spans
216~arcsec at 8.35 GHz, with a radio-jet
that crosses the northern lobe to hit into a bright hotspot. 
There is little evidence of a southern jet, but
a twin hotspot in the southern lobe is also present
(Baum et al. 1988, Leathy et al. 1997, Hardcastle et al. 1998).

Broad band imaging of the host of 3C~98 reveals a smooth
and slightly elongated elliptical galaxy 
located in a sparse environment (Baum et al. 1988,
Martel et al. 1999). Deeper images reveal a faint shell
as a sign of a past disturbance
(Smith \& Heckman 1989). However, the rotation curves of 3C~98 
show that the stellar kinematics has negligible rotation 
$< 25$~\univel\ and no anomalies (Smith, Heckman \& Illingworth 1990). 
Although the optical colours of 3C~98 are similar to those of
normal elliptical galaxies (Zibel 1996), 
one should note that the colours are modified by the high
Galactic extiction towards the source, \mbox{$A_V=0.986$}
(Schlegel et al. 1998). The $\Delta 4000$\AA\ and [MgFe] indices
found in this radiogalaxy
are characteristic of old metal-rich populations.

An extended narrow line region with a wealth of structure,
and no particular orientation with respect to the radio-axis, is also 
detected in direct narrow-band images (Baum et al. 1988). The narrow emission
lines detected in the optical spectra correspond to highly ionized plasma
(Baldwin, Phillips \& Terlevich 1981, Saunders et al. 1989, Baum, 
Heckman \& van Breugel 1992).

3C~98 has been detected in X-rays by the Einstein satellite
at a flux level 
$f(\mbox{0.5--3keV}) = 1\times10^{-13}$~\mbox{erg cm$^{-2}$ s$^{-1}$}
or $L_X = 4.2 \times 10^{41}$~\unilum\
(Fabbiano et al. 1984). The source detection
was too weak to look for any extension to a point source.

\subsection{3C~192}

3C~192 has an `X' symmetric double-lobe structure which extends 212~arcsec
 at 8.35~GHz, showing bright
hotspots at the end of the lobes (Hogbom 1979, Baum et al. 1988,
Hardcastle et al. 1998).

 According to Sandage (1972),
3C~192 is a member of a small group of galaxies.
Broad band imaging reveals the host of 3C~192 to be a round elliptical
galaxy with a companion of similar size 70~arcsec away, and no obvious
signs of interaction (Baum et al. 1988, Baum \& Heckman 1988). 
The stellar kinematics shows negligible rotation, $< 30$~\univel,
and no disturbances (Smith et al. 1990). 
The spectral shape of 3C~192
also shows a blue excess with respect to our template elliptical galaxy.

Extended narrow line emission is detected, with structures which are
co-spatial with bridges and cocoons detected in radio-emission 
(Baum et al. 1988).
The narrow emission
lines are highly ionized 
(Baldwin et al. 1981, Saunders et al. 1989, Baum, Heckman \& van Breugel 1992).

The Einstein satellite detected 3C~192 in X-rays  
at a flux level 
$f(\mbox{0.5--3keV}) = 1.1\times10^{-13}$~\mbox{erg cm$^{-2}$ s$^{-1}$}, 
or $1.8 \times 10^{42}$~\unilum. The source is 
extended at a 97\% confidence level, $0.8^{+1.7}_{-0.3}$~arcmin,
but a background object might be contaminating the map
(Fabbiano et al. 1984).

\subsection{3C~218 or Hydra A}

3C~218 is one of the most luminous radiosources in the local ($z<0.1$)
Universe, only surpassed by Cygnus-A. Although the radio-luminosity
of 3C~218 exceeds by an order of magnitude the characteristic 
FR~I/FR~II break luminosity, it has an edge-darkened FR~I
morphology
(Ekers \& Simkin 1983, Baum et al. 1988, Taylor et al.
1990). The total size of the radio structure extends for about 7~arcmin,
such that the radio-jets, which flare at 5 arcsec, are curved and
display 'S' symetry.

3C~218 has been optically identified with the cD2 galaxy Hydra~A
(Simkin 1979),
which dominates the poor cluster Abell 780. This however is an X-ray bright
cluster with  $L_{\mbox{\scriptsize X}} \approx 2 \times 10^{44}$~\unilum\
in the 0.5 -- 4.5 keV range, 
as seen by the Einstein satellite (David et al. 1990).
The total bolometric luminosity has been estimated 
to be $5\times10^{44}$~\unilum\ from 0.4--2~keV ROSAT observations 
(Peres et al. 1998).
The thermal model that best fits the data supports the existence
of a cooling flow which is depositing mass in the central regions of the
cluster at a rate of  $264^{+81}_{-60}$\Msun\ yr$^{-1}$. 

Hydra~A has an associated type II cooling flow nebula (Heckman et al. 1989), 
characterized by 
high \Ha\ and X-ray luminosities, but
relatively weak [N~II] and [S~II] and strong [O~I]~\ldo{6300} emission
lines, usually found in LINERs. The \Ha\ extended narrow line emission
(Baum et al. 1988) actually fills the gap between the radiolobes.

The \ldo{2200}, \ldo{2700}, $B$-band, $B-V$  and, also, the $U-I$  
continuum colours of the center of
Hydra~A have been attributed to a $\sim$~10 Myr burst of star formation
involving $10^8$ to $10^9$\Msun\ 
(Hansen, J\o rgensen \& N\o rgaard-Nielsen 
1995, McNamara 1995). This is further supported by the detection 
of strong absorption
lines of the Balmer series in the near-UV spectrum of the nucleus
(Hansen et al. 1995, Melnick et al. 1997).
We also find strong absorption Balmer lines,
which we identify as originating in
blue supergiant or giant B stars. One of our best two matches
in the $D\lambda_1$ classification we use in this work, B3I stars,
also indicates ages 7 to 8 Myr.

Heckman et al. (1985) found that the stellar kinematics has negligible
rotation ($13 \pm 18$~\univel), but their observations were limited 
to the \ldo{4200 --5700} region, and did not include the higher
Balmer lines in absorption. On the other hand, Ekers \& Simkin (1983)
report very fast rotating stars in the central 20~kpc of the radiogalaxy.
A two dimensional analysis of the blue spectrum shows a tilt of
the Balmer absorption lines of
$450 \pm 130$~\univel\ in the central 3~arcsec, while the Ca~II H,K lines 
do not show any displacement (Melnick et al. 1997). This tilt 
is further confirmed by our data. The conclusion derived by
Melnick et al. (1997) is that the young stars have formed a disk which 
is rotating perpendicular to the position of the radio-axis. The 
star-formation disk
has also been detected in $U$-band images (McNamara 1995).

\subsection{3C~285}

The host galaxy of 3C~285 has been identified with the brightest member 
of a group of galaxies
(Sandage 1972). Optical imaging of the galaxy reveals an elliptical main 
body and a distorted 
S shape envelope aligned with a companion galaxy
$\sim 40$~arcsec to the northwest (Heckman et al. 1986).
Narrow band imaging shows that the S-shaped  
extension is due to continuum 
emitting structures (Heckman et al. 1986, Baum et al. 1988).
The narrow emission lines are originated by photoionization 
with a high ionization parameter (Saunders et al. 1989, Baum et al. 1992).

Sandage (1972) found that the $B-V$ colour of 3C~285 is much bluer
than that of a normal elliptical galaxy. Our observations show that the blue 
light
of the nucleus (inner 2~arcsec) is dominated by a burst which contains A2I
stars, and thus has an age of 10 to 12~Myr.

Saslaw, Tyson and Crane (1978) identified a bright blue slightly resolved
object halfway between the nucleus and the eastern radio-lobe, which they
denoted 3C~285/09.6. Optical spectra and imaging obtained by van Breugel 
\& Dey (1993), showed that the knot is at the same redshift as the 
galaxy, and its $UBV$ colours and 4000\AA\ break
are consistent with a burst of 70~Myr, which they interpreted as 
induced by the radio-jet.

3C~285 is a classical double-lobed radiogalaxy of 190~arcsec total extension
at 4.86 GHz, with two hotspots and an eastern ridge showing curvature 
roughly along the line to the optical companion (Leahy \& Williams 1984,  
Hardcastle et al. 1998).

The source has not been detected by the Einstein satellite in X-rays,
at a flux level 
$f(\mbox{0.5--3keV}) <1.5\times10^{-13}$~\mbox{erg cm$^{-2}$ s$^{-1}$}
or $L_X = 4.4 \times 10^{42}$~\unilum\ (Fabbiano et al. 1984).

\subsection{3C 382}

3C~382 has a double-lobe structure, with a clear jet in the northern
lobe that ends in a hotspot. A hotspot in the southern lobe is 
also detected, but a counterpart jet is not clear, although a trail of 
low fractional polarization is detected (Black et al. 1992). The total
3.85 GHz size between hotspots is 179~arcsec (Hardcastle et al. 1998).

Optically, the radiosource is identified with a disturbed
 elliptical galaxy dominated 
by a very bright and unresolved nucleus (Mathews, Morgan \& Schmidt 1964, 
Martel et al. 1999), located in a moderately 
rich environment (Longair \& Seldner 1979). 
The optical spectra 
shows a strong continuum and 
prominent broad-lines photoinized by a power-law type of
spectrum 
(Saunders et al. 1989, Tadhunter, Fosbury \& Quinn 
1989). The stellar population of the host galaxy, as we show
in our study, is barely detected in the nuclear regions.

The Einstein satellite detected 3C~382 in X-rays  
at a flux level 
$f(\mbox{0.5--3keV}) = 1.3\times10^{-13}$~\mbox{erg cm$^{-2}$ s$^{-1}$}, 
or $2 \times 10^{44}$~\unilum\ (Fabbiano et al. 1984). The source is resolved
in ROSAT/HRI observations but its interpretation is debateable since
the luminosity is too strong 
for a galaxy environment which 
is only moderately rich (Prieto 2000).

3C~382 is a variable source at X-ray 
(Dower et al. 1980, Barr \& Giommi 1992), 
radio (Strom, Willis \& Willis 1978),  optical and UV frequencies 
(Puschell 1981, Tadhunter, P\'erez \& Fosbury 1986)

\subsection{DA~240}

This is a double-lobed giant radio-galaxy of 34~arcmin angular size 
between hotspots 
and ongoing nuclear activity at 2.8cm  
(Laing et al. 1983, Nilsson et al. 1993, Klein et al. 1994). 

The amplitude of the angular cross-correlation of sources found in optical
plates around the position of the radio source is weak, 
$A_{\mbox{\scriptsize gg}}=0.101\pm0.118$
(Prestage \& Peacock 1988). Abell clusters at the same redshift 
have values $A_{\mbox{\scriptsize gg}} \gsim 0.3$.

The optical spectra shows weak \Hb\ and [Ne III] and [O~III] narrow 
emission lines,
compatible with a higly ionized medium which is obscured. 
The $\Delta 4000$\AA\ and [MgFe] indices
found in this radiogalaxy
are characteristic of old metal-rich populations.

\subsection{4C~73.08}

4C~73.08 is a giant double-lobed radio-galaxy, with 
13~arcmin angular size between hotspots  (Meyer 1979, Nilsson et al. 1993).

The environment of the radiogalaxy is also weak, with
amplitude of the angular cross-correlation of optical galaxies around the radiosource
of  $A_{\mbox{\scriptsize gg}}=0.203 $
(Prestage \& Peacock 1988). 

4C~73.08 shows a high excitation spectrum typical of
narrow line radio galaxies. The colours of the radiogalaxy
and the $\Delta 4000$\AA\ and [MgFe] indices are 
comparable to those of our reference elliptical galaxy.

%%%%%%%%%%%%%%
%%%% Section 6
%%%%%%%%%%%%%%

\ifoldfss
  \section{Conclusions}
\else
  \section[]{Conclusions}
\fi

We have presented spectra of 7 radiogalaxies in the 
\ldo{3350}\ --\ldo{6000} and 
\ldo{7900}\ --\ldo{9400} range. All radiogalaxies show either a 
clear detection or an indication of detection of the
Ca~II~$\lambda\lambda$8494,8542,8662\AA\ triplet in absorption,
and in 6 of them we detected Balmer absorption lines.

   On the basis of the $\Delta 4000$\AA\ break measurements,
we conclude that 4 of these radiogalaxies contain populations which
are typical of normal elliptical galaxies, 2 have populations younger than
a few hundrer Myr, and in one its stellar population cannot be characterized.

   In the two cases with young bursts, Hydra~A and 3C~285,  
we subtracted the bulge population  
using a normal elliptical galaxy as a template in order to characterize better
the young burst. The \ldo{4000}
and Balmer break index measurements indicate that the young population is
dominated by blue giant  and/or blue supergiant stars:
B3I or B5III for Hydra~A, and A2I for 3C~285. The derived
age of the burst is beween 7 and 40~Myr for Hydra~A, and
10 to 12~Myr for 3C~285. 

  The CaT strength, invoked to support the detection of
young stellar populations in active galaxies, fails to provide a
clear conclusion on the nature of the stars that 
dominate the red light in these radiogalaxies. The CaT could either 
be due to 
the red giant stars that dominate old bulge populations, or
to the red dwarfs of a young starburst ($t \lsim 7$Myr), or 
the red giants and supergiants of a post-starburst ($t\gsim 30$~Myr), 
or a combination of a bulge population and a recent burst of star formation.
A mixed population is again favoured as the interpretation of the red spectra.

It is known that although the hosts of FR~II sources look
like ellipticals, 
few of them have true elliptical galaxy properties: magnitudes, colours, 
and structural parameters show a wider dispersion than in normal ellipticals
(Baum et al. 1988, 
Zirbel 1996). Most of the radiogalaxies in our sample have reported
structural
disturbances in their optical morphologies, show signs of interactions, 
have close companions, belong to rich environments and/or
have signatures of cooling flows. These are phenomena that facilitate 
carrying  large quantities of gas to the centers of the galaxies
and can power the AGN and/or provoke  bursts of
star formation.

Good quality data in the blue region of this sample is necessary 
in order to constrain the ages of the young populations involved,
especially in the cases of  3C~98, 3C~192, DA~210, 3C~382 and 4C~73.08,
where our bulge subtractions led to poor signal-to-noise and therefore 
unreliable results.

A detailed analysis of the ages of the last burst of star formation will
set the relative chronology of the onset of
the radio and starburst activity in these galaxies, and shed new light
into the relationships between jets, AGN and star formation.

\section*{Acknowledgments}
This work has been supported in part by the 
`Formation and Evolution of
Galaxies' Network set up by the European Commission under contract 
ERB FMRX-CT96-086 of its TMR programme. We thank PATT for awarding observing 
time. IA, ET and RJT also thank the 
Guillermo Haro Programme for Advanced Astrophysics of INAOE for the 
oportunity it 
gave us to meet and make progress on the project during the 1998
workshop 'The Formation and Evolution of Galaxies'. 
GC acknowledges a PPARC Postdoctoral Research Fellowship, and ET an IBERDROLA
Visiting Professorship to the Universidad Aut\'onoma de Madrid.
We thank J. Gorgas for providing  the CaT spectra of the sample
of comparison elliptical galaxies prior to publication, M. 
Garc\'{\i}a-Vargas for suggestions on how to improve the fringing removal,
and an anonymous referee for crucial comments on the relevance of Balmer
indices in old populations.

\begin{table*}
 \begin{minipage}{140mm}
\begin{center}
\caption{Journal of spectroscopic observations} \label{obs}
\begin{tabular}{lcccccrrcr}
\hline
  object & $P_{{\mbox{\scriptsize 178MHz}}}$ & $z$ & $V$ & type & date 
& PA & exp. & grating & $l$(1 arcsec) \\
	 & $\times 10^{25}$~W~Hz$^{-1}$~Sr$^{-1}$	  &     & mag &      &
& $^o$ & s  &  & pc\\
\hline

3C ~98	&  1.5  &  0.0306  &  15.9  & RG & 1998 Feb. 19 & 221 & 8290 & 
R300B & 422\\
	&	&	   &	    &	 &	        &     & 8304 &
R316R & \\

3C ~192	&  2.6  &  0.0598  &  16.1  & RG & 1998 Feb 19 & 133 & 7200 & 
R300B & 786 \\
	&	&	   &	    &	 &	        &     & 7200 &
R316R & \\

3C ~218	(Hydra A)&  22.0  &  0.0533  &  12.6  & RG & 1998 Feb 20 & 113 & 10800 & 
R300B & 708 \\
      &	&	   &	    &	 &	        &     & 10800 &
R316R & \\

3C ~285	&  2.5  &  0.0794  &  15.9  & RG & 1998 Feb 20 &  66 & 10800 & 
R300B & 1010\\
        &	&	   &	    &	 &	        &     &  7200 &
R316R & \\

3C ~382	&  2.2  &  0.0578  &  15.4  & RG & 1998 Feb 19 &  47 & 6200 & 
R300B & 762 \\
	&	&	   &	    &	 &	        &     & 6200 &
R316R & \\

4C ~73.08&  1.6 &  0.0581  &  16.0  & RG & 1998 Feb 19 &  68 & 7200 & 
R300B & 765 \\
	&	&	   &	    &	 &	        &     & 7200 &
R316R & \\

DA 240	&  0.9  &  0.0350  &  15.1  & RG & 1997 Nov 7  &  61 & 3600 & 
R600B & 479\\
	&	&	   &	    &	 &	        &     & 3900 &
R600R & \\

NGC~4374&  ---  &  0.0034  &  10.1     & E  & 1998 Feb 19 & 135 & 1000 & 
R300B & \\
	&	&	   &	    &	 &	        &     & 1000 &
R316R & 49\\

G191-B2B & ---   &  ---	   &       & S  & 1997 Nov 7   & ---  & 300  &
R600B & \\
	&	&	   &	    &	 &	        &     & 300 &
R600R & \\
 	 &	   &  	   &       &    & 1997 Nov 8   &    & 300  &
R600B & \\
	&	&	   &	    &	 &	        &     & 300 &
R600R & \\

HR~1908 & ---   &  ---	   &        & S  & 1998 Feb 20 & --- & 2  &
R300B & \\
	&	&	   &	    &	 &	        &     & 3 &
R316R & \\

HR~4575 & ---   &  ---	   &       & S  & 1998 Feb 19 & ---  & 1  &
R300B & \\
	&	&	   &	    &	 &	        &     & 1 &
R316R & \\

HR~4672 & ---   &  ---	   &       & S  & 1998 Feb 19 & ---  & 1  &
R300B & \\
	&	&	   &	    &	 &	        &     & 1 &
R316R & \\

HR~5826 & ---   &  ---	   &       & S  & 1998 Feb 19 & ---  & 1  &
R300B & \\
	&	&	   &	    &	 &	        &     & 1 &
R316R & \\

HR~8150 & ---   &  ---	   &       & S  & 1997 Nov 8   & ---  & 1  &
R600B & \\
	&	&	   &	    &	 &	        &     & 1 &
R600R & \\

HZ~44 & ---   &  ---	   &       & S  & 1998 Feb 19   & ---  & 60  &
R300B & \\
	&	&	   &	    &	 &	        &     & 180 &
R316R & \\
        &      &  	   &       &    & 1998 Feb 20   & 	 & 60  &
R300B & \\

\hline
\end{tabular}
\end{center}
\end{minipage}
\end{table*}

\begin{table*}
% \begin{minipage}{140mm}
\begin{center}
\caption{Calcium triplet index and velocity dispersions } \label{EW}
\begin{tabular}{lccc}
\hline
  object & $\sigma$ & CaT$_{\mbox{\scriptsize u}}$ & CaT \\
         &  km/s    & \AA     & \AA  \\        
\hline

3C~98	 &  250     &  5.9     & 6.5\\  

3C~192   &  241     &  5.4    & 6.0\\

Hydra~A	 &  292     &  6.1   & 7.0 \\

3C~285	 &  202     &  $>2.4$  &  --- \\

4C~73.08 & ---	    &  $>5.2$ 	& ---\\

DA 240	 & 265 	    &  5.5	& 6.2 \\

NGC~4374 & 361	    & 5.3	& 6.8  \\

\hline
\end{tabular}
\end{center}
%\end{minipage}
\end{table*}

\begin{table*}
 \begin{minipage}{140mm}
\begin{center}
\caption{Break indices } \label{break}
\begin{tabular}{lcccc}
\hline

	 & {original spectra} & \multicolumn{3}{c}{bulge-subtracted spectra} \\

object   & $\Delta$4000\AA & $\Delta$4000\AA & $D$ & $\lambda_1$ \\
	 &  mag      &  mag      &      &  \AA   \\
\hline

3C~98	 &  2.1 &    & & 	\\ 
	  	      	               
3C~192   &  1.9	&    & & 	\\
	  	      	               
Hydra~A	 &  1.4	& 1.2  & 0.26--0.28  & 3746--3752  \\
	  	      	               
3C~285	 &  1.6	& 1.0  & 0.48--0.53  & 3740--3746 \\
	  	      	               
4C~73.08 &  2.2	&    &	& 	\\
	  	      	               
DA 240	 &  2.2	&    &	& 	\\
	  	      	               
NGC~4374 &  2.3	&    &	& 	\\

\hline
\end{tabular}
\end{center}
\end{minipage}
\end{table*}

\begin{table*}
 \begin{minipage}{140mm}
\begin{center}
\caption{Lick indices} \label{lick}
\begin{tabular}{lccccccccccc}
\hline
  object & Balmer & Mg b$_{\rm u}$ & Mg b & Fe5270$_{\rm u}$ & Fe5270 &
Fe5335$_{\rm u}$ & Fe5335 & [MgFe] & Mg$_{2\rm u}$ & Mg$_{2}$  \\
         &  \AA    & \AA     &  \AA   &  \AA   &   \AA  &  \AA & \AA & \AA & mag   &  mag   \\  
\hline			       			         	       
			       			         	       
3C~98	 &  3.5     &  4.76  & 5.03 &  2.73  & 3.06 & 2.66 & 3.16  & 3.95   & 0.26   & 0.28  \\  
			       			         	       
3C~192   &  $>3.6$     &  5.48  &  5.48 & 2.52 & 2.52 & 2.50   & 2.50  & 3.70   & 0.27   & 0.29  \\
			       			         	       
Hydra~A	 &  3.5     &  4.78  & 5.60 & 1.66 & 2.48 & 1.48  & 2.80 & 
3.84   & 0.20   & 0.22  \\
			       			         	       
3C~285	 &  4.4     &  3.37  & 3.51 & 1.67 & 1.85 & 1.50 & 1.76   & 
2.52   & 0.18   & 0.20  \\
			       			         	       
4C~73.08 &  3.8	    &  4.82  &  5.28 & 2.80  & 3.39 & 2.27 & 3.21  & 
4.17   & 0.23   & 0.25  \\
			       			         	       
DA 240	 &  4.5	    &  4.01  &  4.28   & ---  & ---    & ---  & ---    
& --- & ---    & ---   \\
			       			         	       
NGC~4374 &  1.4     &  4.58  & 4.80 &  2.48  & 2.70 & 2.12   & 2.39 & 
3.49   & 0.30    & 0.32 \\

\hline
\end{tabular}
\end{center}
\end{minipage}
\end{table*}

\include{figures1}

\include{figures2}

\end{document}

%% file: figures1.tex
\newpage
\begin{figure*}   
%    \cidfig{6.0in}{26}{144}{573}{664}{\DIRFIGS sky.ps}
    \caption{Sky line spectrum and atmospheric absorption
 template used to reduce the data}
\end{figure*}

%\newpage
\begin{figure*}   
%    \cidfig{6.0in}{35}{69}{562}{755}{\DIRFIGS 3c98.eps}
    \caption{{\bf (a)} Spectrum of 3C~98 in the observer's frame: 
top pannel is scaled to show the emission 
lines detected in the blue arm spectra; middle pannel shows in greater 
detail the absorption lines in the same wavelength interval; bottom 
pannel shows the CaT region detected in the red arm spectra.}
\end{figure*}

%\newpage
 \setcounter{figure}{1}
\begin{figure*}   
 %  \cidfig{6.0in}{35}{69}{562}{755}{\DIRFIGS 3c192.eps}
    \caption{{\bf (b)}Spectrum of 3C~192} 
\end{figure*}

%\newpage
 \setcounter{figure}{1}
\begin{figure*}   
%    \cidfig{6.0in}{35}{69}{562}{755}{\DIRFIGS 3c218.eps}
    \caption{{\bf (c)}Spectrum of Hydra~A: the red arm spectrum is not flux calibrated} 
\end{figure*}

%\newpage
 \setcounter{figure}{1}
\begin{figure*}   
%   \cidfig{6.0in}{35}{69}{562}{755}{\DIRFIGS 3c285.eps}
    \caption{{\bf (d)}Spectrum of 3C~285: the red arm spectrum is not flux calibrated}
\end{figure*}

%\newpage
 \setcounter{figure}{1}
\begin{figure*}   
%    \cidfig{6.0in}{35}{69}{562}{755}{\DIRFIGS 3c382.eps}
    \caption{{\bf (e)}Spectrum of 3C~382}
\end{figure*}

%\newpage
 \setcounter{figure}{1}
\begin{figure*}   
%    \cidfig{6.0in}{35}{69}{562}{755}{\DIRFIGS 4c73.eps}
    \caption{{\bf (f)}Spectrum of 4C~73.08}
\end{figure*}

%\newpage
 \setcounter{figure}{1}
\begin{figure*}   
%    \cidfig{6.0in}{35}{69}{562}{755}{\DIRFIGS da240.ps}
    \caption{{\bf (g)}Spectrum of DA~240}
\end{figure*}

%\newpage
 \setcounter{figure}{1}
\begin{figure*}   
%   \cidfig{6.0in}{1}{1}{554}{554}{\DIRFIGS n4374.eps}
    \caption{{\bf (h)}Spectrum of NGC 4374}
\end{figure*}

%% file: figures2.tex
\newpage
\begin{figure*}   
%    \cidfig{6.0in}{20}{145}{550}{660}{\DIRFIGS subs2.ps}
    \caption{Bulge-subtraction for (a) Hydra~A and (b) 3C~285.
We show the original spectra of the radiogalaxy (RG), the scaled template 
spectra of the bulge population (E), and the resulting 
bulge-subtracted spectra (subs).}
\end{figure*}

%\newpage
\begin{figure*}   
%    \cidfig{6.0in}{1}{1}{555}{555}{\DIRFIGS conv.eps}
    \caption{Balmer-break measurements of an A2I star: (a) 
original spectrum (b) spectrum convolved with a gaussian filter
to mimic the FWHM of the Balmer lines in the radiogalaxies
($\approx 12.5$\AA). The solid-lines trace the continua and pseudocontinua
of the spectra, and the dashed-lines depict the measurements of the
corresponding $D\lambda_1$ values.}
\end{figure*}

%\newpage
\begin{figure*}   
%    \cidfig{6.0in}{1}{1}{555}{555}{\DIRFIGS rgdfinal.eps}
    \caption{Balmer break measurements of the bulge-subtracted spectra 
of (a) Hydra~A and (b) 3C~285.
The solid-lines trace a wide range of allowed continua and pseudocontinua
of the spectra, and the dashed-lines their corresponding $D\lambda_1$ values.}
\end{figure*}

\begin{figure*}   
%    \cidfig{6.0in}{21}{145}{600}{420}{\DIRFIGS HyA.ps}
    \caption{Estimated absorption profiles of low Balmer lines (thick line), 
compared with the actually observed profiles  in Hydra~A (thin line).}
\end{figure*}

%\newpage
\begin{figure*}   
%    \cidfig{6.0in}{20}{145}{585}{710}{\DIRFIGS Ham.ps}
    \caption{$\Delta 4000$\AA\ break measurements for the radiogalaxies in
our sample (RG) and for elliptical (E), spiral (S) and irregular  (Irr) 
galaxies in the atlas of Kennicutt (1992).
}
\end{figure*}

\newpage
\begin{figure*}   
%    \cidfig{6.0in}{1}{1}{565}{530}{\DIRFIGS bch3.eps}
    \caption{Barbier and Chalonge index plane. The location of the range
of values for 3C~285 and Hydra~A is indicated by solid squares and 
connecting lines. We have also plot  the indices measured
in stars from the library of Jacoby et al. (1984), that have been 
broadened to mimic the width of the Balmer lines in the radiogalaxies,
represented by their corresponding spectral type and luminosity class
({\it e.g.} A2I). The grid of solid lines traces the original locus of 
unbroadened stars. The big symbols at the edges of the grid represent the 
correspondence into spectral classes and luminosity classes of the frame
defined by the grid.
}
\end{figure*}

%\newpage
\begin{figure*}   
%    \cidfig{6.0in}{20}{145}{550}{660}{\DIRFIGS ewRG.ps}
    \caption{CaT index measured for the radiogalaxies (RG),
and compared with the values found in Seyfert~1 (Sy1), Seyfert~2 (Sy2),
LINERs, starburst galaxies (SB), elliptical (E) and spiral and/or lenticular 
(S/S0) galaxies from the samples of Terlevich et al. (1990) and 
Gorgas et al. (priv. communication).
}
\end{figure*}